\documentclass[aps,prl,reprint,groupedaddress]{revtex4-1}

\usepackage{graphicx}
\usepackage{dcolumn}
\usepackage{bm}
\usepackage{amsmath}
\usepackage{amssymb}
\usepackage[section]{placeins}
\usepackage{xcolor}
\usepackage{tikz}

\newcommand{\be}{\begin{equation}}
\newcommand{\ee}{\end{equation}}
\newcommand{\ba}{\begin{align}}
\newcommand{\ea}{\end{align}}

\begin{document}

\title{Universal long-wavelength nonlinear optical response of noble gases}

\author{M. Kolesik}
\email[]{kolesik@acms.arizona.edu}
\affiliation{College of Optical Sciences, University of Arizona, Tucson AZ 85721}

\author{E.~M. Wright}
\affiliation{College of Optical Sciences, University of Arizona, Tucson AZ 85721}

\date{\today}

\begin{abstract}
We demonstrate numerically that the long-wavelength nonlinear dipole moment and
ionization rate versus electric field strength $F$ for different noble gases can
be scaled onto each other, revealing universal functions that characterize the
form of the nonlinear response.  We elucidate the physical origin of the
universality by using a metastable state analysis of the light-atom interaction
in combination with a scaling analysis.  Our results also provide a powerful
new means of characterizing the nonlinear response in the mid-infrared and
long-wave infrared for optical filamentation studies.

\end{abstract}

\pacs{}

\maketitle

\noindent
{\bf Introduction:}  The concept of universality appears across a wide range of
physical systems and reflects the fact that certain features of a system
transcend specific details, either experimental or numerical. Examples abound
and include statistical mechanics for which a large class of systems display
detail-independent properties~\cite{univKadanoff,univText}, the energy and
inverse-energy cascades that appear in three- and two-dimensional turbulence
and give rise to distinct scaling laws, and the universal Efimov
physics~\cite{EfimovPhys} that appears in three-body quantum problems.

The goal of this Letter is to introduce and elucidate a universality in the
long wavelength nonlinear optical response of noble gases.  Here long
wavelength means that the associated photon energies are much less than the
atomic ionization potentials, which implies wavelengths of $2-3~\mu$m or
greater for the noble gases: In this limit the nonlinear optical response
depends only on the instantaneous electric field strength $F$. More specifically,
we demonstrate numerically that the nonlinear dipole moment $p_\text{nl}(F)$ and
ionization rate $\Gamma(F)$ for different gases can be mapped onto each other
using only two species-specific scaling parameters, revealing universal
functions that characterize the form of the nonlinear response.  The numerics
employ previously published single-active electron (SAE) potentials for the
various atoms~\cite{SAETong,SAEkr,SAExe}, and we provide parameterizations of
the universal functions in the Supplemental Online Information (SOI) along with
the two species-specific parameters for each atom.  From an applied perspective
this universality can already be expected to provide a powerful new means of
characterizing the nonlinear response in the mid-infrared (MIR) and long-wave
infrared (LWIR) regions that are currently attracting a great deal of interest
for long distance optical filamentation~\cite{ParisNP,Schuh:17} amongst other
studies. More fundamentally there is the issue of the physical origin of the
observed universality. It has long been recognized that there is a universality
to the field induced ionization rate due to tunneling ionization involving
factors of the form~\cite{Keldysh}
\begin{equation}\label{factors}
\Gamma(F)\propto e^{-2(2U_a)^{3/2}/3F} ,
\end{equation}
with $U_a$ the ionization potential, these factors reflecting the
non-perturbative nature of strong-field ionization.  What is new here is the
universality displayed in the corresponding nonlinear dipole moment.  Here we
use a metastable state analysis~\cite{kolesik:14,ssMESAnoble,cssMESAnoble} to
highlight the intimate relation between the nonlinear dipole moment and
ionization rate in the long wavelength limit, in combination with a scaling
analysis that elucidates the physical origin of the universality.  We conclude
with a summary of our results along with a discussion of some of their
consequences and suggestions for future work.       

\noindent
{\bf Formulation:}
We start from the Schr\"odinger equation for the wave function $\psi({\bf r},t)$
in atomic units [a.u.] for an atom in the presence of an applied electric field
$F(t)$ that is polarized along the x-axis
\begin{equation}\label{SE}
i{\partial\psi\over\partial t} = -{1\over 2} \nabla_{{\bf r}}^2\psi +V_a({\bf r})
  \psi -xF(t) \psi  ,
\end{equation}
where $V_a({\bf r})$ is the atomic potential in the single-active-electron
approximation~\cite{SAETong,SAEkr,SAExe}, and the last term accounts for the
light-atom interaction in the dipole approximation.  The applied electric field
renders the above Schr\"odinger equation almost an open system by virtue of the
inevitable leakage of the wave function away from the atomic core resulting in
ionization.  In order to model what is effectively loss of electrons from the
vicinity of the atomic core, we impose an outgoing or Siegert boundary
condition~\cite{gamow,siegert,hamonou12} on the wave function at large radii.
These outgoing-wave boundary conditions turn the Schr\"odinger operator into a
Non-Hermitian (NH) system~\cite{nHQM} with complex-valued discrete energy
spectrum (we summarize the technicalities of NH theory in the Supplement).
The atomic Schr\"odinger equation may be written in the alternative form
$i{\partial\psi\over\partial t} ={\delta H[\psi]\over\delta \psi}$ with
the Hamiltonian functional
\begin{equation}\label{H}
  H[\psi] = {1\over 2} \int_{ }^{ } d^3{\bf r}~ \psi \left [ -{1\over 2}
    \nabla_{{\bf r}}^2 +V_a({\bf r}) -xF \right ] \psi ,
\end{equation}
which although different from the usual Hermitian result is in keeping with the
ideas of biorthogonal quantum mechanics~\cite{brody} applicable to NH systems
and according to which the wave function is normalized as
$\int d^3{\bf r}~\psi^2({\bf r},t)\equiv\int dzdy \int_Cdx~\psi^2({\bf r},t)=1$,
where the integration along the $x$-axis
proceeds along a contour in the complex plane~\cite{kolesik:14,JeffAMP}
as explained in the Supplement.

\noindent
{\bf Adiabatic approximation:}   Here we consider the long-wavelength limit
meaning that the center frequency of the applied electric field will be much
smaller than the ionization potential $U_a$ of the atom.  Within this limit we
have previously shown that the adiabatic approximation is valid, and the atomic
wave function can be usefully expanded in terms of the instantaneous metastable
states~\cite{Lindl93}, also termed resonant, Gamow~\cite{gamow,Civitarese}, or
Siegert~\cite{siegert,tolstikhin} states, of the Schr\"odinger Eq. (\ref{SE})
for a given $F$ obtained from 
\begin{equation}
  E_j(F)u_j({\bf r},F) = \left [ -{1\over 2} \nabla_{{\bf r}}^2 +V_a({\bf r}) -xF
    \right ]u_j ({\bf r},F) .
\end{equation}
Here $E_j(F)\equiv E_j(F(t))$ is the energy eigenvalue for the $j^{th}$
metastable state and $u_j({\bf r},F)\equiv u_j({\bf r},F(t))$ is the
corresponding eigenstate with normalization $\int d^3{\bf r}~u_j^2({\bf r},F)=1$.
The energy eigenvalues are complex by virtue of the NH nature of the problem,
the instantaneous ground state labeled by $j=0$ having the lowest loss due to
the imaginary part of the energy.  The Metastable Electronic State Approximation
(MESA) \cite{kolesik:14} is founded upon using the metastable states as a basis
for the atomic dynamics, and within the long-wavelength limit we consider
single-state MESA (ssMESA) \cite{ssMESAnoble} in which the atomic wave function
is further assumed to adiabatically follow the instantaneous ground state
\begin{equation}\label{AA}
  \psi({\bf r},t)\approx  u_0({\bf r},F(t))
  \exp \left ( -i\int_0^t dt ' E_0(F(t'))\right )  ,
\end{equation}
where loss requires $\Im(E_0(t))<0$.  Using Eq. (\ref{H}) we note that the
ground state energy may be expressed as $E_0=2H[u_0]$.  We remark that MESA
provides an accurate non-perturbative analysis atom-field interaction
and has been previously verified through comparisons with TDSE
simulations~\cite{cssMESAnoble} and exact results~\cite{kolesik:14}, and was
tested~\cite{SSIvsMESA} against experiments in noble gases.
The single-state method~\cite{ssMESAnoble} is applicable in the adiabatic,
long-wavelength limit considered here, in contrast to the
Kramers-Henneberger~\cite{KHvalidity} and Floquet theory-based
models~\cite{FloquetIon} of strong-field interactions that is better suited
to the short-wavelength limit.

\noindent
{\bf Nonlinear optical response:} 
In the long-wavelength limit considered here the nonlinear properties may be
assessed using ssMESA~\cite{ssMESAnoble} without the need for the correction
terms described in Refs.\cite{cssMESAnoble,Yudin2001}.  In particular, using
the wave function in Eq. (\ref{AA}) the complex dipole moment is calculated
using
\begin{equation}\label{P}
p(F)=\int d^3{\bf r}~u_0({\bf r},F)xu_0({\bf r},F) ,
\end{equation}
and the nonlinear dipole moment is obtained from its real part
$p_\text{nl}(F) = \Re\{p(F) - \lim_{s\to 0}s^{-1} p(s F)\}$, with the subtracted
term being the linear contribution to the dipole moment.   Furthermore, the
decay rate of the instantaneous ground state probability implied by
Eq. (\ref{AA}), which in ssMESA is identified with the ionization rate, is
given by
\begin{equation}\label{Gamma}
\Gamma(F)=-2\Im\{E_0(F)\}  .
\end{equation}
The intimate relation between the nonlinear optical properties may now be
exposed by realizing that the dipole moment may also be expressed as
$p(F)=-2{\partial E_0\over \partial F}$: This is a NH extension~\cite{JeffAMP}
of the well known result for Hermitian systems, here allowing for complex
energies.  Using this expression both the nonlinear dipole moment and ionization
rate can be expressed in a way that exposes their common origin in the complex
ground state energy $E_0(F)$, which is valid in the long-wavelength limit. 

\noindent
{\bf Universal functions \& numerical data:}
We have numerically determined the nonlinear dipole moment $p_\text{nl}(F)$ and
the ionization rate $\Gamma(F)$ as functions of the field strength $F$ for the
noble gas atoms labeled
by $a=He, Ne, Ar, Kr, Xe$ \cite{ssMESAnoble,cssMESAnoble}.  The maximum value
for the field strength $F=0.2$ is chosen larger than those arising during
optical filamentation $F\approx 0.05$ (peak intensities around
$10^{14}$ W/cm$^2$).  For the numerics we employ previously calculated and
tabulated single-active electron (SAE) potentials $V_a({\bf r})$ for the
various atoms~\cite{SAETong,SAEkr,SAExe}, these having ionization potentials
$U_a$.  The outgoing Siegert condition on the instantaneous ground state for a
given $F$ is implemented by complexifying the spatial coordinates at large
radii, as described in detail in Ref.\cite{Bahl:14}. 

The main finding from our numerical study is that the nonlinear optical response
can be characterized by the universal forms
\begin{equation}\label{eqn:scaling}
 p_\text{nl}^{(a)}(F)=\alpha_a^3 {\cal F}^3 M({\cal F}), \quad
\Gamma^{(a)}(F)=\alpha_a G({\cal F})  ,
\end{equation}
with two species-specific scaling parameters $\alpha_a$ and $\beta_a$, and
scaled field strength ${\cal F}=\beta_a F$.  That is, both the nonlinear dipole
moment and the ionization rate of different noble gas atoms can be
simultaneously collapsed, after suitable scaling, onto their respective
universal functions given by $M({\cal F})$ and $G({\cal F})$.   The tabulated
data for these universal functions and their parameterization are made available
in the Supplementary Online Information (SOI) along with the numerically
determined scale parameters $\alpha_a$ and $\beta_a$ for each species. 

\begin{figure}
    \includegraphics[width=0.9\linewidth,clip]{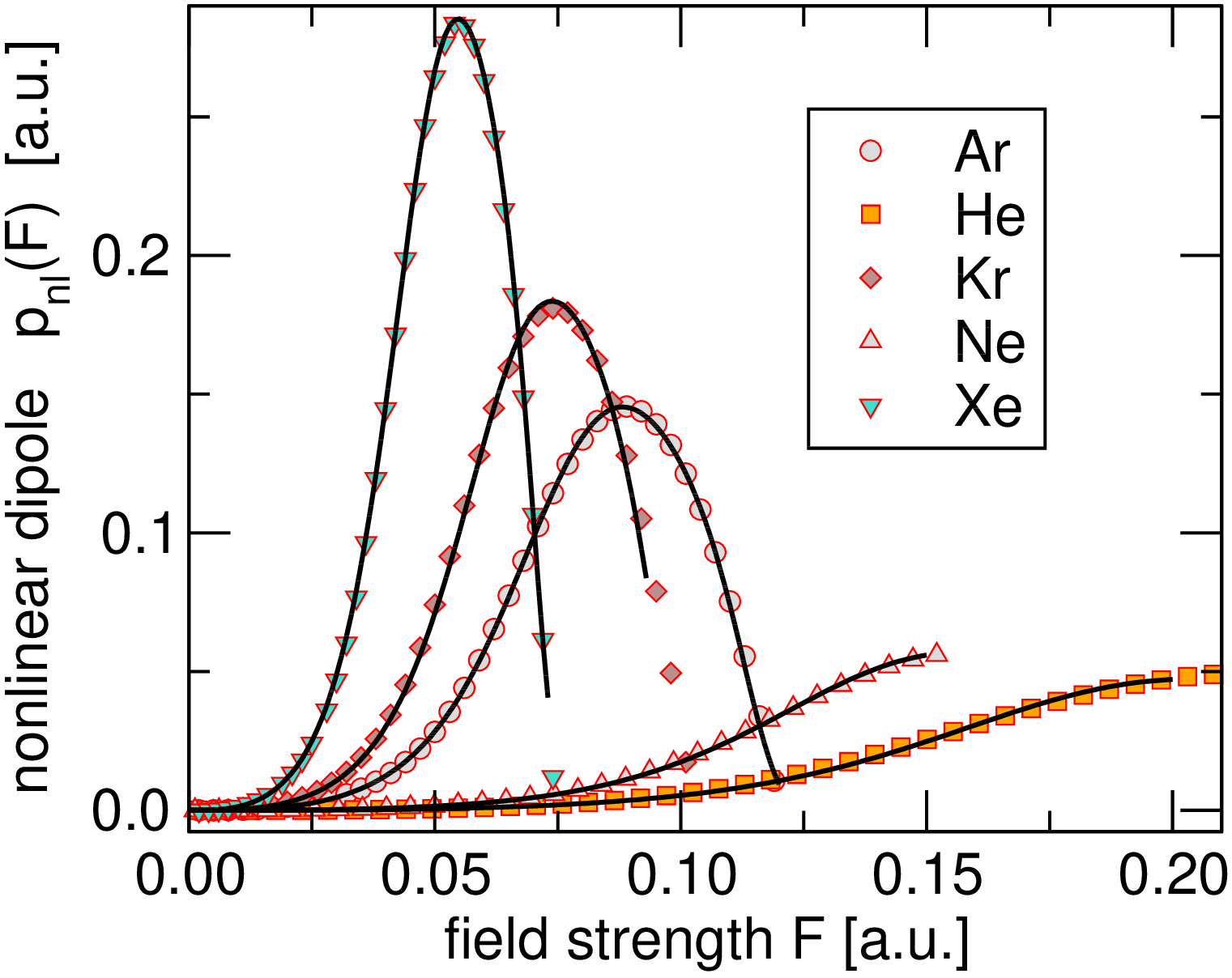}
  \caption{\label{fig:nldipole} (Color online.)
    nonlinear dipole moment $p_\text{nl}(F)$ versus field strength $F$ for
    different noble gas atoms. The symbols represent the ssMESA numerical
    results, while the continuous
    lines are generated using the universal form $M({\cal F})$ in
    (\ref{eqn:scaling}) along with the tabulated values for the scale
    parameters $\alpha_a$ and $\beta_a$ for each species given in the SOI.
  }
\end{figure}

\begin{figure}
    \includegraphics[width=0.9\linewidth,clip]{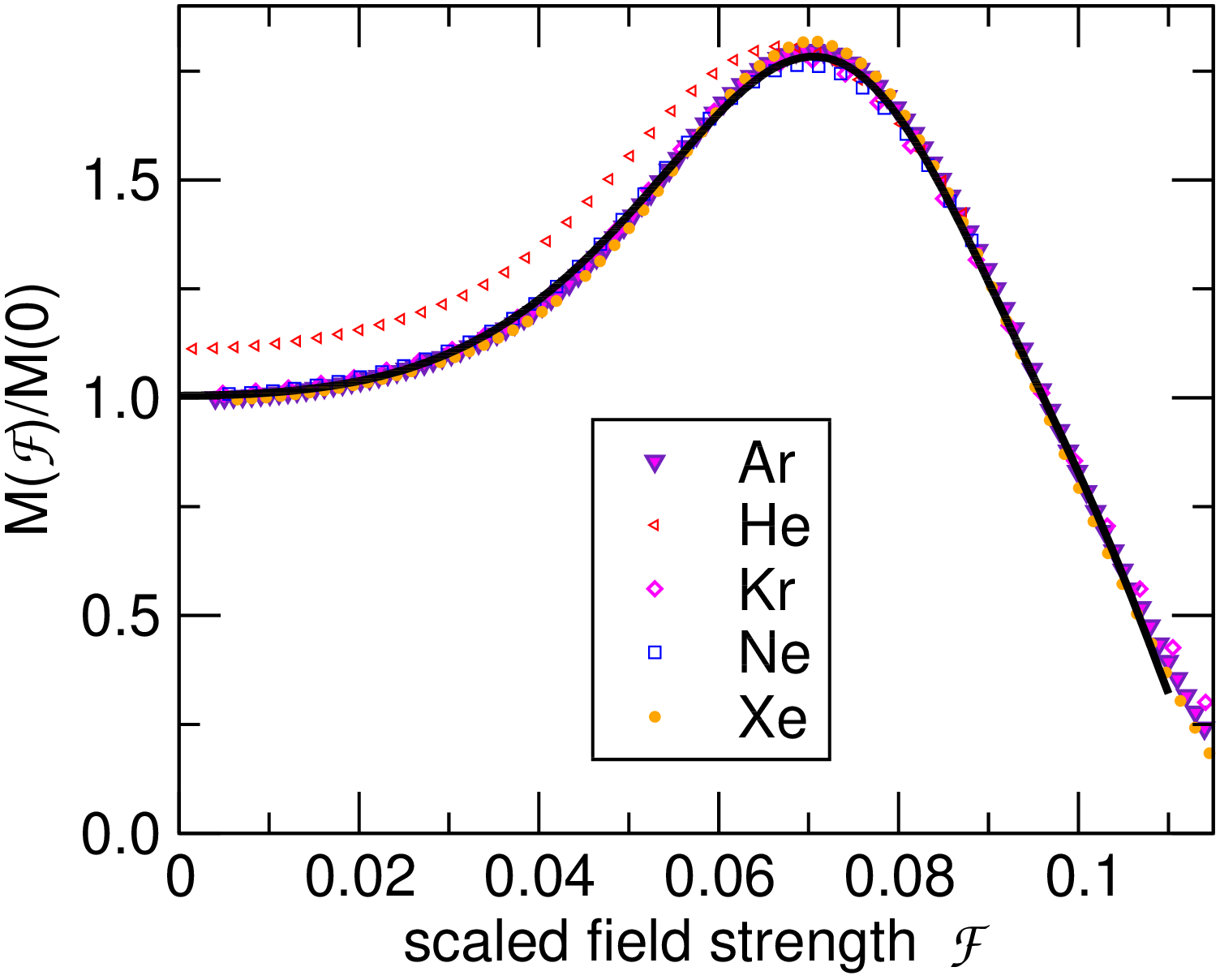}%
  \caption{\label{fig:collapse}  (Color online.)
    Collapse of the nonlinear dipole moment curves in noble gases onto a single
    universal form $M({\cal F})/M(0)$ versus scaled field strength ${\cal F}$
    shown as the thick solid line, where ${\cal F}=\beta_a F$.  The numerical
    data based on ssMESA for each atomic species is shown by the various symbols.
  }
\end{figure}

\begin{figure}
    \includegraphics[width=0.9\linewidth,clip]{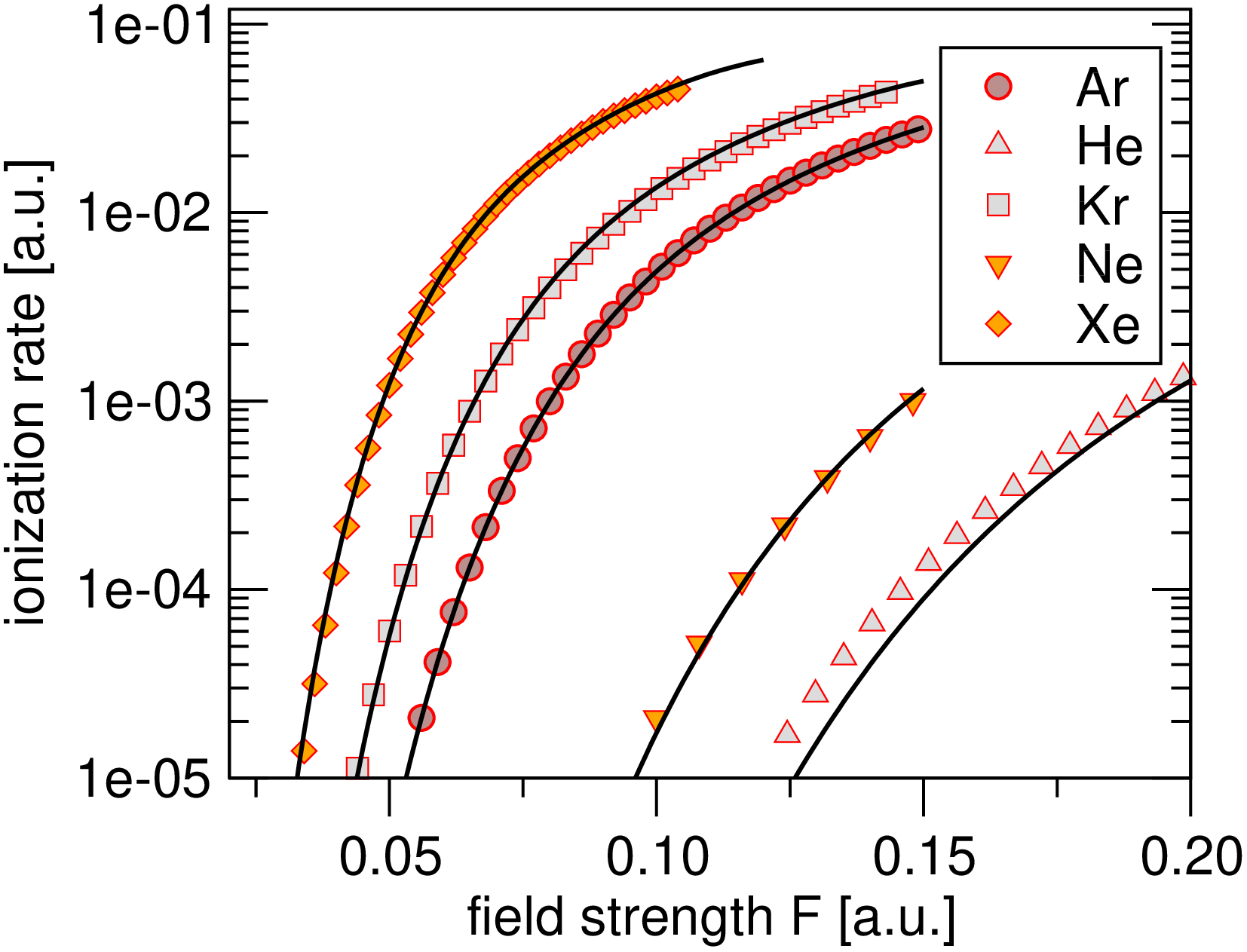}%
  \caption{\label{fig:irateLOG} (Color online.)
    Ionization rate $\Gamma(F)$ versus field strength $F$ for the different
    noble gases. The symbols represent the ssMESA numerical results, while the
    continuous lines are generated using the universal form $G({\cal F})$ in
    (\ref{eqn:scaling}) along with the tabulated values for the scale parameters
    $\alpha_a$ and $\beta_a$ for each species given in the SOI.
  }
\end{figure}

To lay bare the proposed universality Fig.~\ref{fig:nldipole}(a) shows the
nonlinear dipole moment $p_\text{nl}(F)$ versus $F$, the symbols being the
numerical data and the solid lines being the results based on the scaling law
in Eq. (\ref{eqn:scaling}), the quality of the overall fit being evident.
Figure~\ref{fig:collapse}(b) shows the scaled universal function
$M({\cal F})/M(0)$ versus the scaled field strength ${\cal F}$ (thick solid
line), with the numerical data for each species shown by the various symbols.
Here we see excellent agreement with only a few percent deviation between the
case of He and the other species at lower field strengths, a point we shall
return to later.  Such deviations are compatible with the numerical accuracy of
the MESA calculations: Nevertheless, it is clear that the noble gases share to
a significant degree a common functional shape in their nonlinear dipole moment.
The corresponding results to Fig.~\ref{fig:nldipole}(a) for the ionization rate
are shown in Fig.~\ref{fig:irateLOG}, the quality of the fit based on the
universal function $G({\cal F})$ being equally evident, and for both low and
high field strengths.  
  
The numerical data in conjunction with the universal functions scaling laws
(\ref{eqn:scaling}) clearly demonstrate that once $M({\cal F})$ and
$G({\cal F})$ are determined, two parameters are sufficient to characterize the
long-wavelength nonlinear optical response, and this is a main message of this
Letter.  Furthermore, we have performed parallel numerics for the nonlinear
optical properties using different published models for the SAE potentials and
find that although the scale parameters $\alpha_a$ and $\beta_a$ may vary, the
universal functions $M({\cal F})$ and $G({\cal F})$ remain remarkably robust,
an example of this being given in the SOI.  This further bolsters our claim that
the nonlinear optical response of the noble gases is universal: That is, the
functional form of the nonlinear optical properties versus field strength are
more robust than the absolute magnitude of these properties obtained for a given
SAE potential. 

\noindent
{\bf Origin of the universality:}   In the text surrounding Eq. (\ref{Gamma})
we established the intimate relation between the nonlinear dipole moment
$p_\text{nl}(F)$ and the ionization rate $\Gamma(F)$ via their mutual dependence
on the instantaneous ground state energy $E_0(F)$:  This provides a basis for
why they can both simultaneously display universal behavior in the
long-wavelength limit.  Without loss of generality we therefore proceed by
concentrating on the ionization rate as a route to understanding the
universality.

Physically, in the long-wavelength limit considered here, field-induced
ionization arises from tunneling of the bound electron in the atomic core
region through the saddle point region in the composite potential $(V_a-Fx)$
formed by the SAE potential and the applied field.  Since all the noble gas SAE
potentials are to a large degree Coulomb-like for displacements beyond a few
atomic units, it is reasonable that the spatial structure of the saddle point
should be similar for the different noble gases: This in turn would lead to
similar accelerating wave function tails past the saddle point, with concomitant
similar tunnel currents.  We contend that this similarity physically underpins
the universality, and below we develop a scaling argument for the similarity of
the composite potentials in the vicinity of the saddle point.

To proceed we first express Eq. (\ref{factors}) for the ionization rate, which
reflects the universality alluded to, as
$\Gamma({\cal F})\propto e^{-2(2)^{3/2}/3{\cal F}}$, from which we identify
${\cal F}=U_a^{-3/2}F=\beta_a F$.  This simple argument provides an excellent
first approximation to the species-dependent coefficient $\beta_a=U_a^{-3/2}$
that we determined numerically by fine tuning by a few percent around this value,
see the SOI for details.

In the next step we examine the scaling properties of the time-independent
Schr\"odinger equation for the instantaneous ground state
\begin{equation}
E_0 u_0 = \left[ -\frac{1}{2} \nabla_{{\bf r}}^2 + V_a({\bf r})-F x \right]u_0\ .
\end{equation}
Motivated by the above discussion we use the scaled field strength
${\cal F}=s_a^3 F$ with scale factor $s_a=U_a^{-1/2} =\beta_a^{1/3}$.  Then if we
concomitantly scale the spatial coordinate as ${\bf R}={\bf r}/s_a$, the ground
state energy as ${\cal E}_0=s_a^2 E_0=E_0/U_a$, and the atomic potential as
$V_a({\bf r})\rightarrow {1\over s_a}V_a(s_a{\bf R})$, we obtain the scaled
ground state Schr\"odinger equation
\begin{equation}\label{scaledSE}
  {\cal E}_0 u_0 = \left [ -\frac{1}{2}\nabla_{{\bf R}}^2 + s_a V_a(s_a{\bf R}) -
    {\cal F}X  \right  ]u_0\ .
\end{equation}
The key point is that if the scaled atomic potential $s_aV_a(s_a {\bf R})$ were
strictly scale invariant with respect to $s_a$, as it is for a purely Coulomb
potential, then the scaled ground state Schr\"odinger equation could be solved
once and the results for the different species could be obtained using the
scaling above: In this case strict universality would be present.  Furthermore,
based on this analysis the dipole moment, obtained as the expectation value of
the displacement $x$, should scale as $s_a=U_a^{-1/2}$ and this provides a
first-order estimate of the scale parameter $\alpha_a= U_a^{-1/2}$.  

\begin{figure}
    \includegraphics[width=0.9\linewidth,clip]{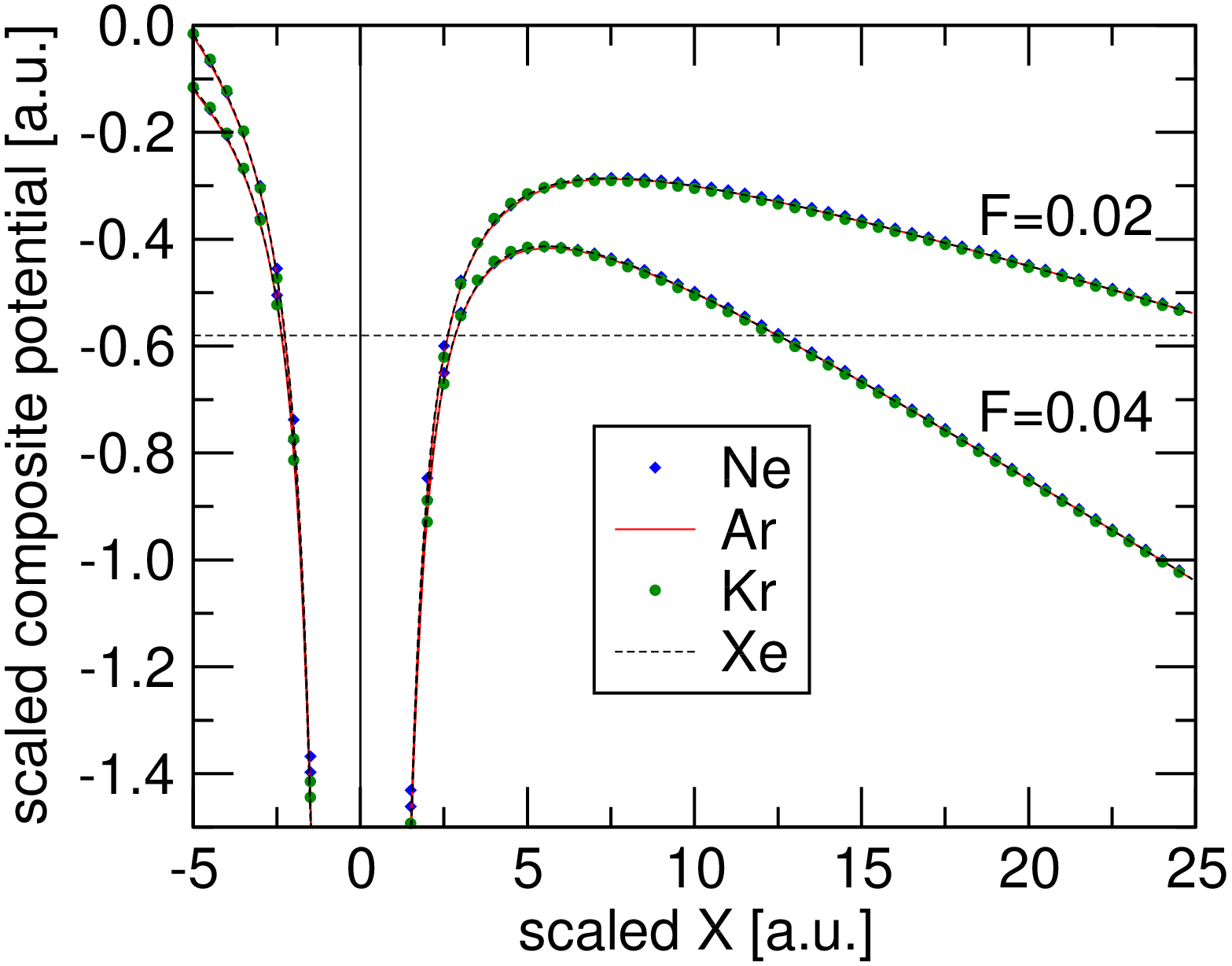}%
  \caption{\label{fig:scaledVs} (Color online)
    Scaled total composite potentials versus scaled coordinate $X$ for the four
    noble gas species in the vicinity of the saddle point of the potential.}
\end{figure}

The SAE potentials for the noble gas atoms do not, however, strictly display
this scale invariance, and this begs the question of why the observed
universality appears for the nonlinear optical properties?  To address this
issue Fig.~\ref{fig:scaledVs} shows the scaled composite potentials,
$(s_aV_a(s_a X)-{\cal F}X)$, versus scaled coordinate $X=s_a x$ for the various
species, and for two values of the scaled field strength ${\cal F}=0.02, 0.04$.
What is key is that in a broad spatial range around the saddle point the
composite potentials collapse onto a single form to a high degree of
approximation.  Since the saddle point dictates the tunneling properties, and
the scale invariance applies in its spatial proximity, this may be identified
as the source of the observed scaling rules and associated universality in the
ionization rate for the noble gases, and by extension also the universality of
the nonlinear dipole moment.  

\begin{figure}
    \includegraphics[width=0.9\linewidth,clip]{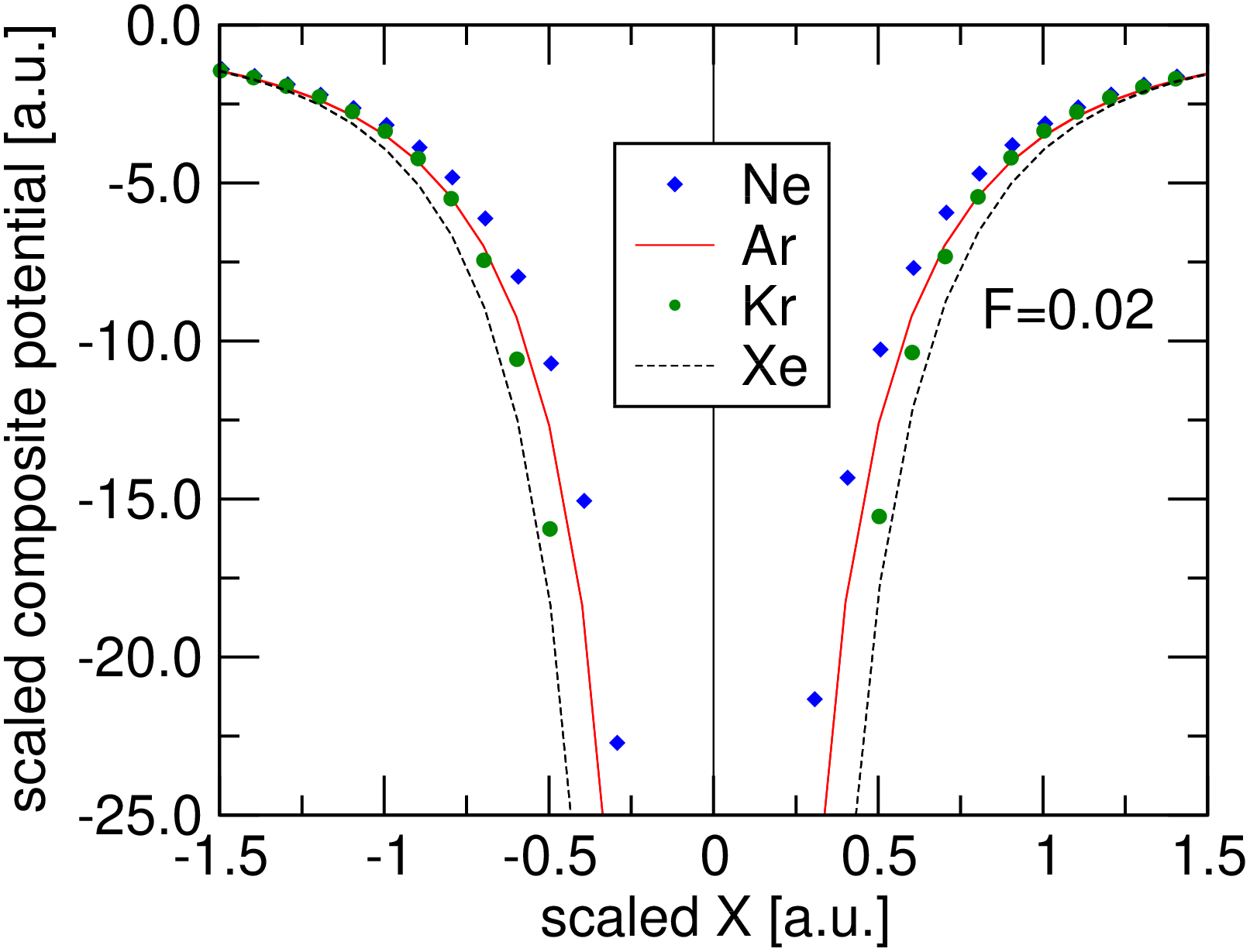}%
  \caption{\label{fig:scaledVsuppl} (Color online) 
    Scaled composite potentials for four noble gas species for small distances
    from the  nucleus. Here, the different species can not be scaled onto each
    other.
  }
\end{figure}

However, it must be acknowledged that the scale invariance does not apply
everywhere for the SAE potentials, as illustrated in Fig.~\ref{fig:scaledVsuppl}
which is the same as Fig.~\ref{fig:scaledVs} for a smaller range of scaled
scaled displacements. In particular, for small distances from the nucleus, the
scaled potentials of different species start to deviate from each other
significantly.  However, the ground state wave functions of the relevant states
in Ne, Ar, Kr, and Xe have a zero at the origin by virtue of the fact that their
field-free ground state is a p-like orbital:  This may underpin why the detailed
structure of the SAE potentials close to the nucleus have a relatively small
effect on the nonlinear optical properties and their universality.   This also
suggests why the results for He showed the largest deviations, though amounting
to only a few percent, since its field-free ground state is a s-like orbital
which is more concentrated around the atomic nucleus.  We speculate that this
might be the reason we see that the properties of He do not scale as well as the
other four gases, and it is possible that He belongs to a different
``universality class.''

\noindent
    {\bf Summary and conclusions:}  In summary, using the Metastable Electronic
    State Approach (MESA) we have demonstrated numerically that in the
    long-wavelength limit the nonlinear optical properties of the noble gases
    can be captured in scaling functions, or master curves, for the nonlinear
    dipole moment and ionization rate versus field strength: This covers
    wavelengths of $2-3$ microns or greater depending on the species, and also
    field strengths that encompass those arising in optical filamentation.
    Using a scaling argument we have traced the physical origin of the
    universality reflected in the scaling functions as arising from the fact
    that the saddle point in the composite potential, both atomic potential
    plus dipole interaction, is almost invariant between the different species:
    Since the ionization rate is dictated mainly by the saddle point structure,
    and we established an intimate link between the nonlinear dipole and
    ionization rate, the universality follows. 

Our results are of both applied and fundamental significance. On the fundamental
side, here we find that whilst the scaling functions are remarkably robust the
scaling parameters are much more dependent on the details of the SAE potentials
of the various atomic species.  This is characteristic of other physical systems
displaying universality, where robust scaling laws appear but with exponents and
absolute magnitudes that depend on the system details, in our case the core
structure of the SAE potentials.  This suggest that deeper ideas associated with
universality, such as renormalization group or universality classes, may be at
play here.   On the applied side, it is now the case that with the scaling
functions established the long-wavelength nonlinear optical properties for a
given species can be obtained using only two scaling parameters.  This is of
significance given current efforts to extend optical filamentation studies into
the MIR and LWIR regimes.  Whilst numerics based on specific SAE potentials can
offer some idea of the values of the scale parameters, experiments will be the
ultimate arbiter of their values, and we hope our work will motivate such
experiments.  Once the scale parameters are obtained our work provides a new
approach to a) the characterization of the nonlinear optical properties for
propagation studies, and b) the interpretation of experimental data.

\begin{acknowledgments}

  This material is based upon work supported by the Air Force Office of
  Scientific Research under award numbers FA9550-18-1-0183 and FA9550-16-1-0121.

\end{acknowledgments}

\newpage

\setcounter{equation}{0}
\setcounter{figure}{0}
\setcounter{table}{0}
\setcounter{page}{1}
\makeatletter
\renewcommand{\theequation}{S\arabic{equation}}
\renewcommand{\thefigure}{S\arabic{figure}}
\renewcommand{\bibnumfmt}[1]{[S#1]}
\renewcommand{\citenumfont}[1]{S#1}

\begin{widetext}
\begin{center}
\textbf{\large Supplemental Material:\\Universal long-wavelength nonlinear optical response
  of noble gases}
\end{center}
\end{widetext}

\section{Master curves}

\vspace{-2mm}

We express the scaling properties of the imaginary part of
the resonant energy $\Gamma$ and of the nonlinear dipole moment $p_\text{nl}$  
as functions of the external field strength $F$ in the form
\begin{align}\label{eqn:scaling}
\Gamma^{(a)}(F)\!&=\!\alpha_a G( \beta_a F )  \nonumber \\
p_\text{nl}^{(a)}(F)\!&=\!\alpha_a^3 (\beta_a F)^3 M( \beta_a F  )
\end{align}
where $G({\cal F})$ and $M({\cal F})$ are the scaling functions or
``master curves,'' and $\alpha_a$ and $\beta_a$ are two scaling parameters
specific to each species $a=$He,Ne,Ar,Kr,Xe.

While the exact properties of the scaling functions are not known analytically,
we approximate them with the parameterizations given below. For the scaled
ionization rate $G$, we use
 \begin{align}
G({\cal F}) = \nonumber \\
\exp[&0.1692048194155632 - 0.810669873391612/{\cal F} \nonumber \\
  +&56.391127621143774 {\cal F} -823.1689378085457 {\cal F}^2 \nonumber \\
  +&4152.098445656342 {\cal F}^3 - 7390.473021873485 {\cal F}^4  \nonumber \\
  -&0.7555077122737133 \log{\cal F}]
 \end{align}
 where the first two and the last term are motivated by the functional form of
 tunneling ionization rates, and the remaining higher-order terms adjust the
 functional shape.

For the nonlinear dipole moment, we propose the following representation
 \begin{align}
M({\cal F}) =& \nonumber \\ 
\exp[
&5.049542998716102 + 63.07083179334633 {\cal F}^2   \nonumber \\
  +&50607.81357765684 {\cal F}^4 - 6632204.58068692{\cal F}^6 \nonumber \\
  -&1019206807.5646534 {\cal F}^8 + 184485603638.703 {\cal F}^{10} \nonumber \\
     -&7509132994894.033 {\cal F}^{12}]
 \end{align}
 for the scaled nonlinear index $M$. These forms are applicable up to scaled
 field strength  up to ${\cal F}\approx 0.12$.

 It must be emphasized that these
 expressions are nothing but fitting functions, and the coefficients carry no
 physical meaning: It is folly to attempt to impose a physical interpretation
 to individual terms in these expressions. Moreover, the number of displayed
 digits does not  imply accuracy --- we list them here make it possible to
 precisely reproduce  calculations for this work.

Naturally, there are infinitely many equivalent ways to represent a scaling
function.  Here we choose to take Argon as our ``reference,'' and initially
approximate the  master curves by fitting previously published  numerical
results for this species.  After we estimate the species-specific scaling
parameters, the shape of the master curves  is adjusted via fitting to the
appropriately scaled numerical results for multiple
gas species --- this procedure is described next.

\section{Scaling Parameters}

\vspace{-2mm}

In order to utilize $G$ and $M$ to represent the nonlinear response in a given
gas, two scaling parameters must be specified. The following table lists
the calculated values for $\alpha_a$ and $\beta_a$. 

\medskip

\centerline{{\bf Table I:} Nonlinear Response Scaling Parameters}
 \begin{tabular}{|l|l|l|}
   \hline
     Species & $\alpha_a$ & $\beta_a$         \\ 
   \hline
     Helium       & 0.6867429501537934  & 0.43699038577877725 \\
     Neon         & 0.7312465607920158  & 0.5706469503314686 \\
     Argon        & 1.008406125021463,  & 1.0015305349341797 \\
     Krypton      & 1.0875887749796427  & 1.2091505108407845 \\
     Xenon        &1.25918930244883     & 1.6064710433471137 \\
   \hline
 \end{tabular}

\medskip
 
We have obtained these values by matching simultaneously $p_\text{nl}^{(a)}(F)$
and $\Gamma^{(a)}(F)$ to the numerical values from \cite{cssMESAnoble} over
a range of field strengths $F$ between zero and a field strength at which
the nonlinear dipole curve of the species saturates and starts to decrease.
Since we adopt Argon as our reference, we introduce the scaled scaling
parameter $\bar\beta_a =\beta_a/\beta_{Ar}$, and the initial guess for
$\bar\beta_a$ was taken from the numerical values of the ionization potential
relative to Argon, $\bar\beta_a \sim (U_a/U_\text{Ar})^{-3/2}$. Subsequently,
parameter values were refined interactively (trial and error) by fitting each
species to the master curve. The resulting values were finally utilized to
initiate an iterative procedure in which a master curve was obtained from a fit
to scaled data from Ne,Ar,Kr, and Xe, followed by improvement of the
$\alpha_a$ and $\beta_a$ via fitting to the resulting master curve.
From a comparison of different matching procedures we estimate that the
expected variation of the scaling coefficients is a few percent.

Because our characterization of the scaling invariance of the nonlinear
response is ultimately based on numerical characterization of model atoms, it
is important to ask how do different models of the same species compare from
this standpoint. Figure~S1 shows an example comparison between two different
SAE-based models of Krypton.

In panel a) we plot the numerical results for two versions of the
single-active-electron potentials~\cite{SAEkr}, here shown in the form of the
nonlinear dipole moments and the adiabatic ionization rates. Obviously the
results exhibit a difference albeit not a large one. However, it is easy to
see that the different models still predict the same {\em shape} of these
response curves. This is demonstrated in panel b) where we have
applied  two-parameter scaling in order to collapse the two pairs of curves.

\medskip

\centerline{
\includegraphics[clip,width=0.49\linewidth]{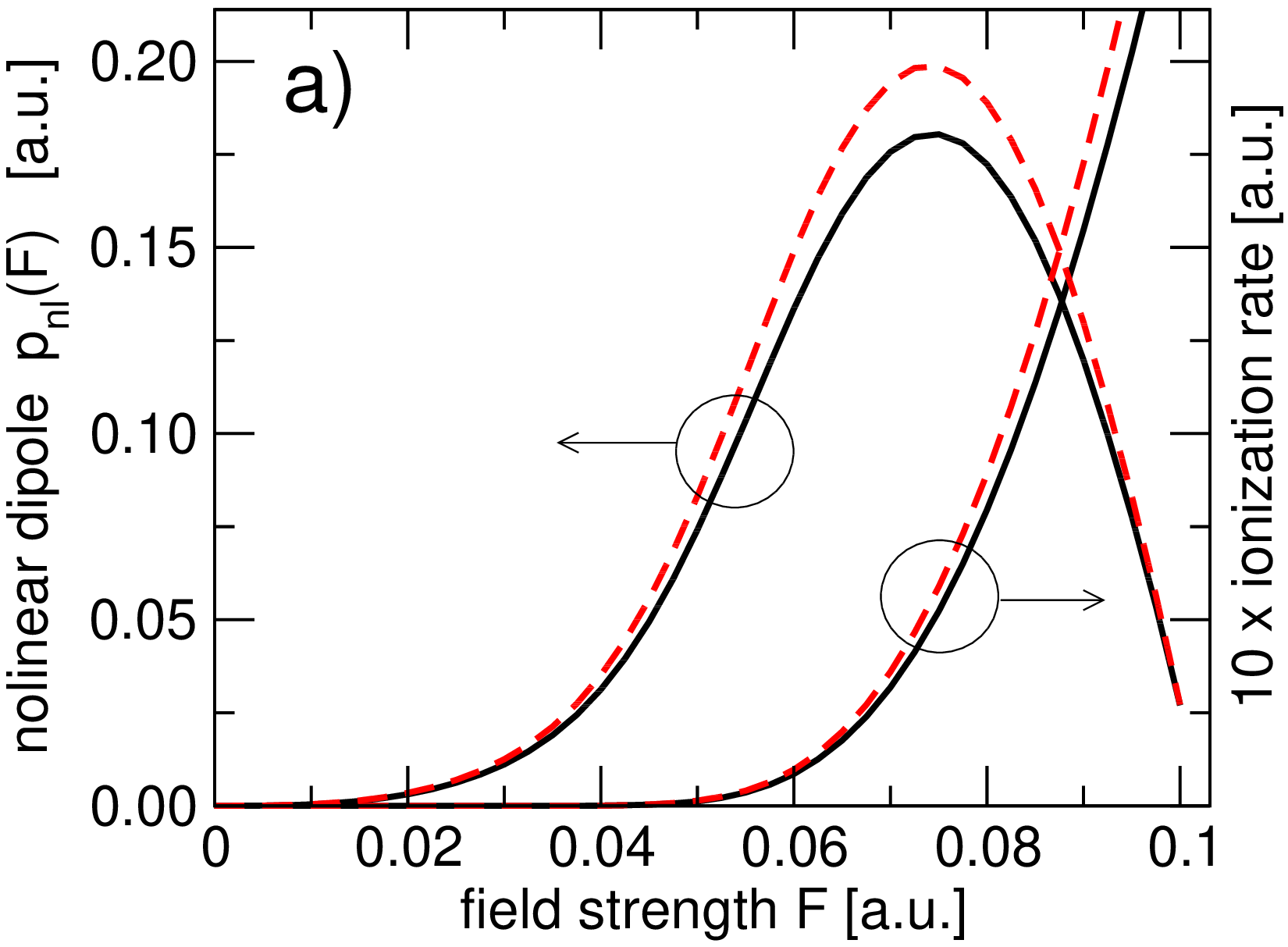}
\includegraphics[clip,width=0.49\linewidth]{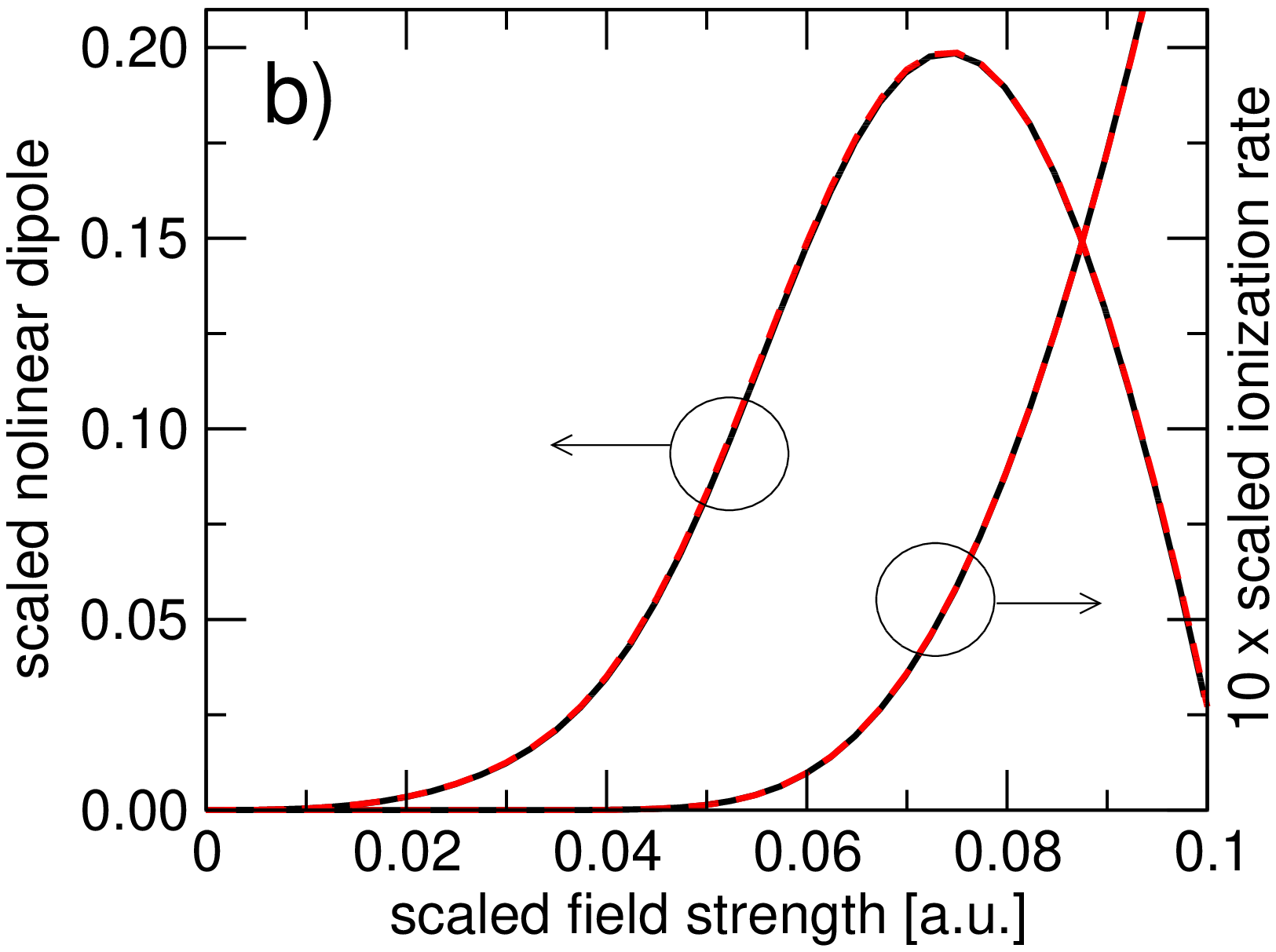}
}
\begin{minipage}{0.9\linewidth}
  \small
  {\bf Figure S1.} (Color online)
  Nonlinear dipole moment (left vertical axes) and ionization rates (right
  vertical axes)   obtained for two different SAE potentials of Krypton are
  shown in panel a).   Panel b) depicts the   same curves after suitable
  horizontal and vertical scaling, demonstrating that the curves share
  the same shape.
  \end{minipage}

\medskip

A similar observation can be made for Argon, where different SAE potentials are
also available and give rise to the same functional shapes of both the
nonlinear dipole and ionization rate.

\section{Metastable Electronic States}

\vspace{-2mm}

Here we summarize some of the ideas from non-Hermitian~\cite{nHQM} or
biorthogonal quantum mechanics\cite{bohm_resonances,brody,garcia-calderon} that
we allude to in the main text.  In the long-wavelength limit the adiabatic
approximation applies, and the nonlinear optical  response  is governed by the
electronic wavefunction ``slaved'' to the external electric field.  Within the
single-active-electron approximation~\cite{SAETong,SAEkr,SAExe}, the standard
(i.e. Hermitian) eigenvalue problem for a system subjected to a homogeneous
field of strength F reads
\begin{equation}
  -{1\over 2} \nabla_{{\bf r}}^2\psi +V_a({\bf r})\psi -x F \psi   = E \psi \ ,
\label{eq:hermit}
\end{equation}
where $V_a$ is the single-active-particle atomic potential, and the domain of
this operator consists of $L_2$-integrable functions. Interestingly, no matter
how weak the field $F$ is, there exist no bound states for this Hamiltonian.
Instead, the spectrum is purely continuous, and it covers the whole real
axis~\cite{bookCycon}. Needless to say, this complicates
both analytic and numerical treatments.

Instead of attempting to extract the nonlinear response properties from a
continuum of energy-eigenstates, we adopt a non-Hermitian
approach~\cite{moiseyev_review} based on the Metastable Electronic State
Approach (MESA)~\cite{kolesik:14}. The advantage of the method is that even a
single wavefunction can provide a very good approximation~\cite{ssMESAnoble}.
Higher-energy states can be approximately included in order to account for
non-adiabatic effects for shorter wavelengths~\cite{cssMESAnoble}. The method
has been tested against exact solutions~\cite{Brown2011,kolesik:14} as well as
against experiments~\cite{SSIvsMESA}.

The metastable wavefunctions utilized in the single-state MESA are the
so-called Stark resonances~\cite{JeffAMP} --- they can be obtained as solutions
to the same equation (\ref{eq:hermit}) but the domain of the operator is
selected by the requirement that at large distances from the nucleus the
wavefunction must behave as an outgoing wave,
$\psi(x\to\infty)\sim\exp(+i k x)$.
These boundary conditions~\cite{gamow,siegert,tolstikhin} make the operator
non-Hermitian
and its spectrum is discrete, with complex-valued energies (see e.g.
\cite{cavalcanti,emmanouilidou,Brown} for exactly solvable one-dimensional
examples). The non-Hermitian eigenstates depend on the field strength $F$, and
as $F\to 0$ some of them approach the bound-states of the original Hermitian
system (one should note that this limit is highly nontrivial).

Importantly, all complex-energy eigenstates are orthogonal in the following
sense:
\begin{equation}
\int\!\! dydz \int_C\!\! dx\ \psi_{E1}(x,y,z) \psi_{E2}(x,y,z) =\delta_{E1,E2} \ ,
\end{equation}
where the integration along $x$ follows a contour in the complex
plane~\cite{Brown,JeffAMP}. This contour takes over the role of the coordinate
axis $x$, and its precise shape is unimportant as long as for large $|x|$ it
deviates from the real axis into the upper complex half-plane (Fig.S2).
The crucial point is that outgoing waves on such a contour decay to zero at
infinity. 

\medskip

\centerline{
\includegraphics[clip,width=0.79\linewidth]{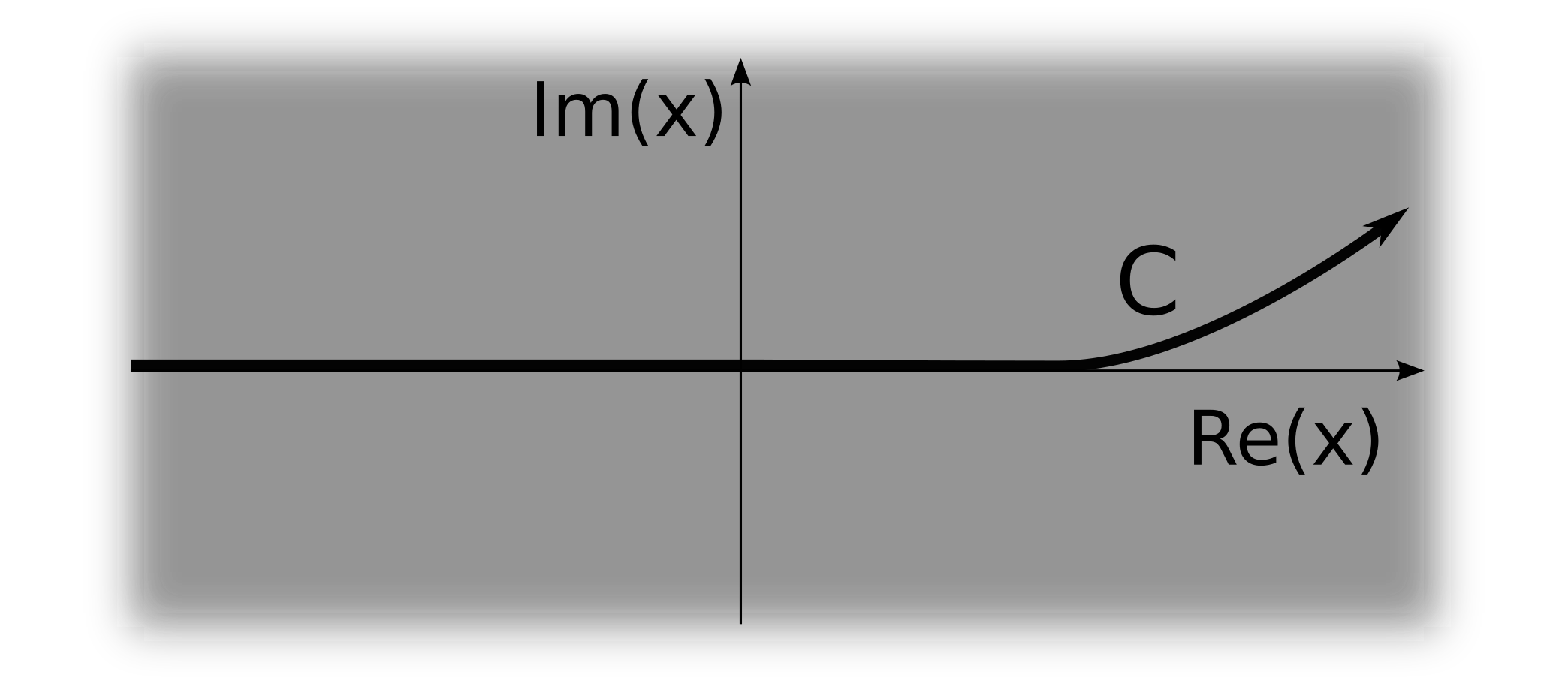}
}
\begin{minipage}{0.9\linewidth}
  \small
  {\bf Figure S2.} (Color online)
  Complex-valued spatial axis: Contour C follows the real axis except at very
  large distance   from the origin, when it starts to deviate into upper
  complex plane. Outgoing Stark resonance   states are normalizable and
  mutually orthogonal when integrated along such a contour.
  \end{minipage}

\medskip

Expectation values for the resonance states~\cite{Civitarese} can be defined in
an analogous way, for example the component of dipole moment along the
direction of the field is calculates as
\begin{equation}
p(F) = \int\!\! dydz\! \int_C\!\!\! dx \ \psi(x,y,z,F)\  x\ \psi(x,y,z,F)  \ ,
\end{equation}
where we made the dependence on the field strength explicit.

It is important to note that other methods exist for regularization of the
integrals involving non-integrable resonance
states~\cite{gyarmati,zeldovich,julve,romo}. In particular, our method
utilizing a complex contour integration is closely related to the so-called
external complex scaling approach~\cite{moiseyev_review} to non-Hermitian
quantum mechanics.

\end{document}